\documentclass[%
 reprint,
superscriptaddress,
 amsmath,amssymb,
 aps,
]{revtex4-2}

\usepackage{graphicx}
\usepackage{dcolumn}
\usepackage{subfigure,color}
\usepackage{bm}
\usepackage{hyperref}
\hypersetup{
	colorlinks=true,
	linkcolor=blue,
	filecolor=blue,      
	urlcolor=blue,
}

\begin{document}

\preprint{APS/123-QED}

\title{Intelligent Metasurfaces with Continuously Tunable Local Surface Impedance for Multiple Reconfigurable Functions}

\author{Fu Liu$^\ddagger$}\email{fu.liu@aalto.fi}
\affiliation{Department of Electronics and Nanoengineering, Aalto University, P.O.~Box 15500, FI-00076 Aalto, Finland}
\thanks{These authors contributed equally to this work.}
\author{Odysseas Tsilipakos$^\ddagger$}\email{otsilipakos@iesl.forth.gr}
\affiliation{Institute of Electronic Structure and Laser, FORTH, GR-71110 Heraklion, Crete, Greece}
\author{Alexandros Pitilakis}
\affiliation{Institute of Electronic Structure and Laser, FORTH, GR-71110 Heraklion, Crete, Greece}
\affiliation{Department of Electrical and Computer Engineering, Aristotle University of Thessaloniki, Thessaloniki GR-54124, Greece}
\author{Anna C. Tasolamprou}
\affiliation{Institute of Electronic Structure and Laser, FORTH, GR-71110 Heraklion, Crete, Greece}
\author{Mohammad Sajjad Mirmoosa}
\affiliation{Department of Electronics and Nanoengineering, Aalto University, P.O.~Box 15500, FI-00076 Aalto, Finland}
\author{Nikolaos V. Kantartzis}
\affiliation{Institute of Electronic Structure and Laser, FORTH, GR-71110 Heraklion, Crete, Greece}
\affiliation{Department of Electrical and Computer Engineering, Aristotle University of Thessaloniki, Thessaloniki GR-54124, Greece}
\author{Do-Hoon Kwon}
\affiliation{Department of Electrical and Computer Engineering, University of Massachusetts Amherst, Amherst, Massachusetts 01003, USA}
\author{Maria Kafesaki}
\affiliation{Institute of Electronic Structure and Laser, FORTH, GR-71110 Heraklion, Crete, Greece}
\affiliation{Department of Materials Science and Technology, University of Crete, GR-71003 Heraklion, Crete, Greece}
\author{Costas M. Soukoulis}
\affiliation{Institute of Electronic Structure and Laser, FORTH, GR-71110 Heraklion, Crete, Greece}
\affiliation{Ames Laboratory---U.S. DOE and Department of Physics and Astronomy, Iowa State University, Ames, Iowa 50011, USA}
\author{Sergei A. Tretyakov}
\affiliation{Department of Electronics and Nanoengineering, Aalto University, P.O.~Box 15500, FI-00076 Aalto, Finland}

\date{\today}

\begin{abstract}
Electromagnetic metasurfaces can be characterized as intelligent if they are able to perform multiple tunable functions, with the desired response being controlled by a computer influencing the individual electromagnetic properties of each metasurface inclusion. In this paper, we present an example of an intelligent metasurface which operates in the reflection mode in the microwave frequency range. We numerically show that without changing the main body of the metasurface we can achieve tunable perfect absorption and tunable anomalous reflection. The tunability features can be implemented using mixed-signal integrated circuits (ICs), which can independently vary both the resistance and reactance, offering complete local control over the complex surface impedance. The ICs are embedded in the unit cells by connecting two metal patches over a thin grounded substrate and the reflection property of the intelligent metasurface can be readily controlled by a computer. Our intelligent metasurface can have significant influence on future space-time modulated metasurfaces and a multitude of applications, such as beam steering, energy harvesting, and communications.

\bigskip\noindent This paper is published at Phys. Rev. Applied. \\
DOI: \href{http://dx.doi.org/10.1103/PhysRevApplied.11.044024}{10.1103/PhysRevApplied.11.044024}
\end{abstract}

\maketitle

\section{Introduction}
Metasurfaces, the two-dimensional versions of metamaterials, have been an intriguing research subject in recent years. This attention results from the many exotic properties that can be attained with these ultrathin, practically two-dimensional structures~\cite{Minovich:2015,Chen:2016,Hsiao:2017}. Engineering the meta-atoms comprising the metasurfaces gives rise to different interactions of the surface with incoming electromagnetic waves. Therefore, we have the ability to control  wave propagation~\cite{EnghetaBook,Soukoulis:2011}, which leads to a vast spectrum of supported applications ranging from wavefront shaping~\cite{Yu:2011,Pfeiffer:2013,Tsilipakos:2018aom,Huang:2017AdvOptMat}, polarization control~\cite{Kruk:2016} and dispersion engineering \cite{Tsilipakos:2018,Dastmalchi:2014} to perfect absorption \cite{Radi:2015prappl,Liu:2017OpEx,Tsilipakos:2017,Isic:2015} and holography \cite{Li:2017NatCom}. The operation frequency of recent demonstrations extends from microwaves to optical frequencies~\cite{Glybovski:2016PhysRep,Shadrivov:2012PRL,Shadrivov:2017bookch9}.

The versatility of metasurfaces is greatly enhanced if they are equipped with a tuning mechanism~\cite{Zheludev:2012NatMat,Turpin:2014IJAP} since, for instance, we can achieve the same functionality but at different frequencies or have the ability to switch the functionalities at a particular working frequency. Different tuning mechanisms result in different levels of function-tuning capabilities. A global tuning approach, where all unit cells are tuned concomitantly by means of an external stimulus, provides control over global functions such as the absorption level, resonance frequency, and polarization state of plane waves \cite{Liu2018,Komar:2017,Bi2015,Kafesaki2012,Seo:2010}. In these cases, the main body of a metasurface is tuned. Besides global tuning, a more compelling mechanism is local tuning, which offers a possibility of individually tuning the properties of each meta-atom (unit cell) \cite{Zhu:2010APL,Zhao:2013NJP,Mias20071955,Tsilipakos:2018aom}. This approach enables us to achieve more exotic functionalities such as beam steering, focusing, imaging, and holography \cite{Liu2018,Cui:2014,Yang:2016SciRep,Li:2017NatCom,Chen2017}.

The existing local tuning strategies have mainly concentrated on a binary phase-state scenario, or a continuous reactance tuning scenario, e.g., by introducing diodes or varactors \cite{Li:2017NatCom,Liu2018,Zhu:2010APL,Zhao:2013NJP,Mias20071955,Cui:2014,Yang:2016SciRep,Chen2017}, respectively. In such scenarios the impedance characteristics (reactive and resistive impedance) of the meta-atom vary only between two different states or follow a line. However, as the property of the meta-atoms is represented by both  reactive and resistive characteristics, the aforementioned scenarios are not exploring the full capability of the tuning area in the impedance space. In general, full control requires independent control over both the reactive and absorptive characteristics of meta-atoms. Actually, we can even move further, assuming that both the reactive and absorptive characteristics can take different states in a continuous way. Then the developed metasurface can be transformed into an intelligent one, if a computer can determine the optimal states of  meta-atoms for realizing the desired functionality. This is the extreme tuning scenario that can utilize the larger range of impedance of the metasurfaces. Hence, such intelligent metasurfaces can provide a multitude of possibilities programmatically controlled by a computer.

In this paper, we introduce a design of an intelligent metasurface that utilizes tunable integrated circuits (ICs) incorporated with the meta-atoms in each unit cell. The control computer enables us to tune both the reactive and absorptive properties of the unit cells, in a concomitant or independent way. We show that without changing the geometrical properties of the proposed metasurface and its materials, we can switch from one function to another at the same frequency due to the regrouping of the unit cells into new supercells. Also, each function can be tuned at different frequencies and incident angles for both transverse-electric (TE) and transverse-magnetic (TM) polarizations of the incident waves. This work paves the way for the next generation of metasurfaces that are computer intelligent and can be programmed.

The paper is organized as follows. In Sec.~\ref{sec:ucdescrip}, we first give the design of the proposed intelligent metasurface and subsequently, in Sec.~\ref{sec:sectpa}, we present the results showing a possibility of obtaining tunable perfect absorption for different incidence angles. In Sec.~\ref{sec:AnomRefl}, we uncover a possibility of realizing tunable perfect anomalous reflection and, finally, Sec.~\ref{sec:concl} concludes the paper.


\section{Design of intelligent metasurface}\label{sec:ucdescrip}
\begin{figure}[t!]\centering
	\includegraphics{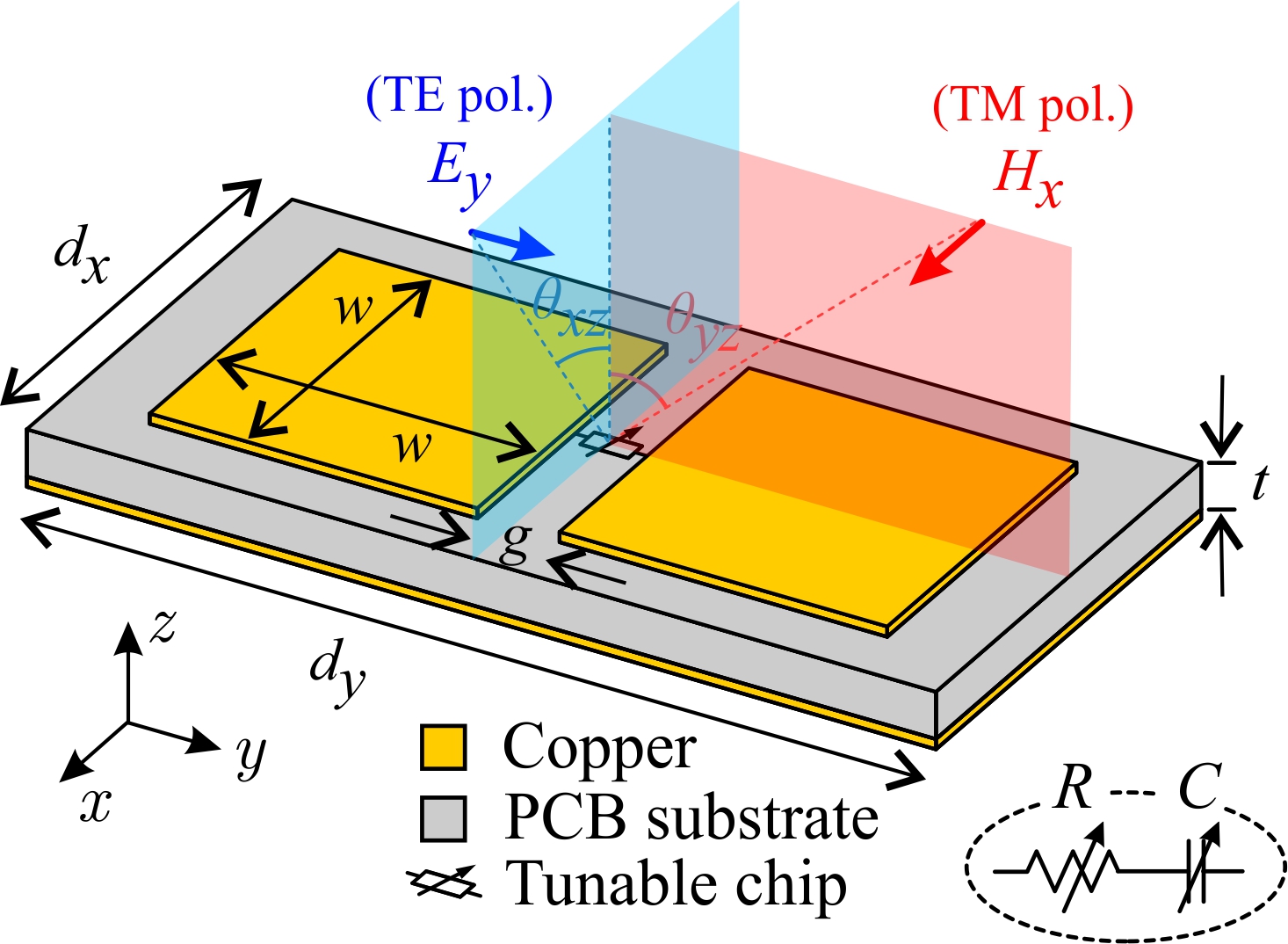}
	\caption{\label{fig:MainSchem} Schematic of the unit cell for the proposed intelligent metasurface. We select the basic topology of a high-impedance surface and find the suitable dimensions for operation in the target frequency range (around 5~GHz) using the analytical formulas \cite{Luukkonen2008}. The dimensions are: $d_x=d_y/2=9.12$~mm, $t=1.016$~mm, $w=8.12$~mm and $g=1$~mm. A tunable ICs chip is incorporated to provide a variable complex impedance and locally modify the surface impedance of the metasurface in a continuous way.}
\end{figure}

The proposed intelligent metasurface topology, a periodical array of metal patches over a continuous ground plane, is known as a microwave-range versatile metasurface, which can be designed to absorb incident waves or shape  the reflected wave fronts~\cite{Sievenpiper1999,Radi:2015prappl,Diaz-Rubio:2017}. Here, we show that this structure can be made tunable and provide software-defined functionalities by equipping unit cells with computer-controllable ICs. 
For demonstration purposes, we design an  intelligent metasurface for multiple tunable functions at 5~GHz, i.e., tunable perfect absorption (TPA) for different incidence angles and tunable anomalous reflection (TAR) toward different directions. In this case, the tunable IC is introduced into the unit cell by connecting two metal patches along the $y$ direction, as schematically shown in Fig.~\ref{fig:MainSchem}. The ICs can be modeled as continuously tunable, lumped complex impedance loads, with a tunable capacitance $C$ (negative reactance) and a tunable resistance $R$. The capacitance typically controls the spectral position of the resonance, while the resistance controls its strength and absorption level (see Figs.~S1 and S2 in Supplemental Material~\cite{SupplementaryInfo} for details). As a result, TPA is a rather straightforward task as when the unit cell is tuned to have resonance at 5 GHz by changing the capacitance $C$, perfect absorption can be achieved by tuning the resistance $R$. Realization of TAR, however, is more demanding. First of all, the unit cell size should be as subwavelength as possible to ensure fine resolution in reflection angles, but it should not be too small to allow practical manufacturing using the available technology. Therefore, we chose $d_x=d_y/2=9.12$ mm (approximately $0.15\lambda$). Secondly, the reflection phase span of the unit cell (in a  periodical setting) should be as wide as possible within the working range of the tunable chip. This requires a high-$Q$ resonance. For our unit-cell configuration, the $Q$ factor is proportional to $\sqrt{C_{\mathrm{eff}}/L_{\mathrm{eff}}}$, where $C_{\mathrm{eff}}$ is the effective capacitance and $L_{\mathrm{eff}}\approx\mu d$ is the effective inductance when the structure is viewed as an equivalent parallel RLC circuit \cite{Luukkonen2008}. The effective capacitance comes from the gaps between patches and the IC contribution, while the effective inductance is due to the grounded dielectric substrate. Therefore, a thinner substrate and larger metal patches (within the practical range) provide better performance (see Fig.~S3 in Supplemental Material for details). Lastly, the design is also constrained by the available $R$ and $C$ values from the ICs.
The finalized unit cell is designed to be fabricated on a thin grounded dielectric substrate with the thickness $t=1.016$~mm (Fig.~\ref{fig:MainSchem}). The square patches have the width $w=8.12$~mm (metallization thickness 17.5~$\mu$m) and are separated from each other by a $1$-mm gap, where the tunable IC is connected.
The thin dielectric substrate is a high-frequency laminate Rogers RT/Duroid 5880 which is characterized by the relative permittivity $\varepsilon_{\rm{r}}=2.2(1-j\tan\delta)$ with the  dissipation factor $\tan\delta=0.0009$. Because of this low dissipation factor, the loss is negligible and the wave does not attenuate noticeably in the substrate. The copper background on the back side of the metasurface prohibits transmission of the incident waves and the performance of the metasurface is characterized by the reflection coefficient. 

Both the tunable resistance $R$ and tunable capacitance $C$ in the ICs can be realized by conventional active electronic circuits using MOSFET technology, e.g., transistors~\cite{AdelS.Sedra2014}. The variable resistance (varistor) can be implemented between the source and drain terminals and its value can be tuned by the gate voltage. For variable capacitance (varactor), one can connect the source and drain terminals. Then the total capacitance between the gate and source terminals is the parallel connection of intrinsic capacitances of the gate-bulk, gate-source and gate-drain terminals, and its value can be  modulated by the gate voltage. Moreover, for both varistors and varactors realized with MOSFET transistors, the required value range can be realized by carefully choosing the channel size of each transistor. In this work, we assume that the tuning range is approximately $0-5~\Omega$ for $R$ and $1-5$~pF for $C$. Therefore, one can control the $RC$ values in each tunable IC (and thus the local surface impedance of each unit cell) by changing the parameters of those active circuits through biasing voltages. This capability allows for local, continuous tunability, greatly enhancing the possible functionalities of the metasurface compared with existing binary encoded~\cite{Cui:2014,Yang:2016SciRep} and tunable varactor~\cite{Zhao:2013NJP} designs.


\section{Tunable Perfect Absorption}\label{sec:sectpa}
\begin{figure}[t!]\centering
	\includegraphics[width=0.4\textwidth]{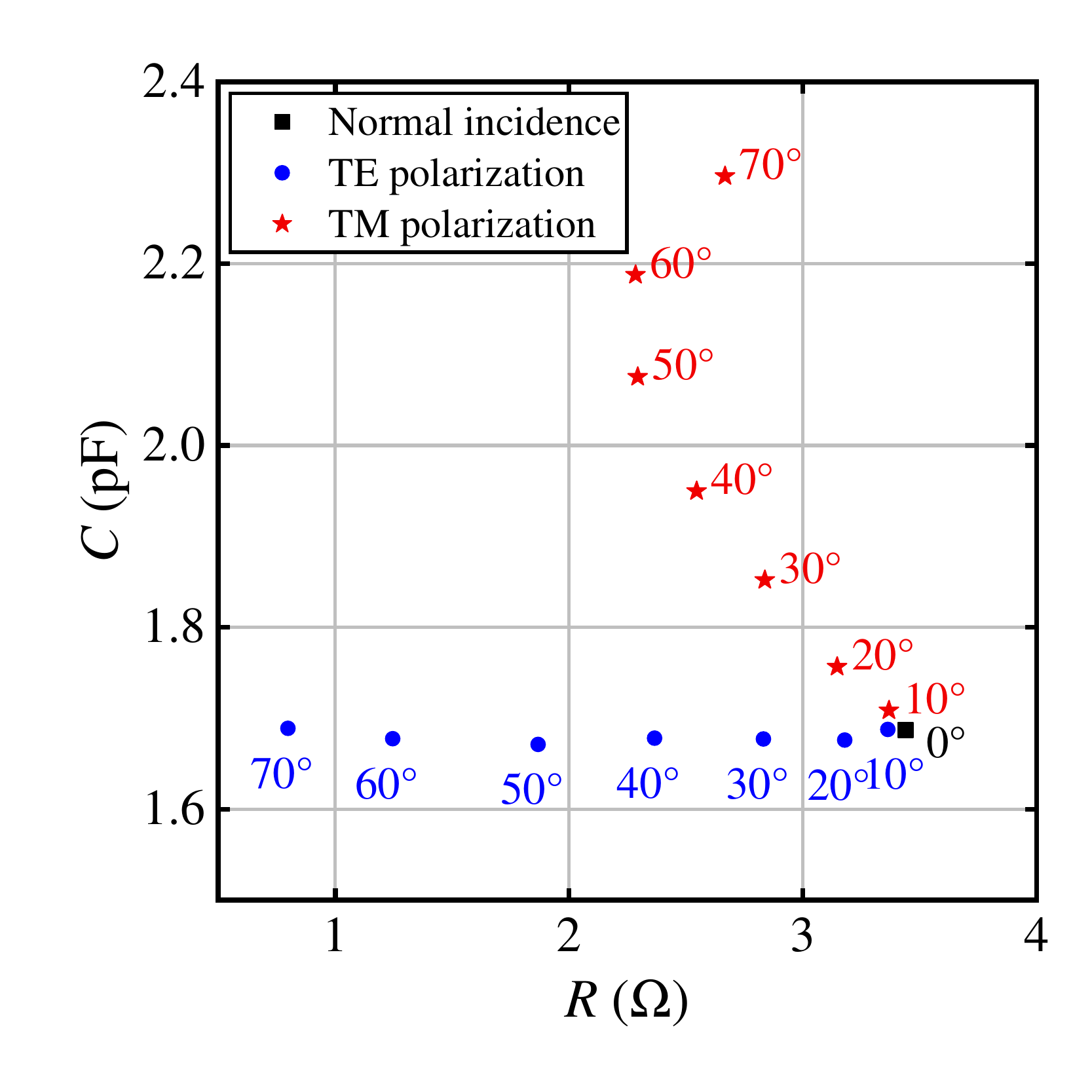}
	\caption{\label{fig:2}The required $R$ and $C$ values of the electronic elements (series topology, see inset in Fig.~\ref{fig:MainSchem}) for achieving perfect absorption at different incidence angles for the two orthogonal polarizations (see Fig.~\ref{fig:MainSchem}). The operation frequency is 5~GHz.}
\end{figure}
\begin{figure*}[!t]\centering
	\includegraphics[width=0.9\textwidth]{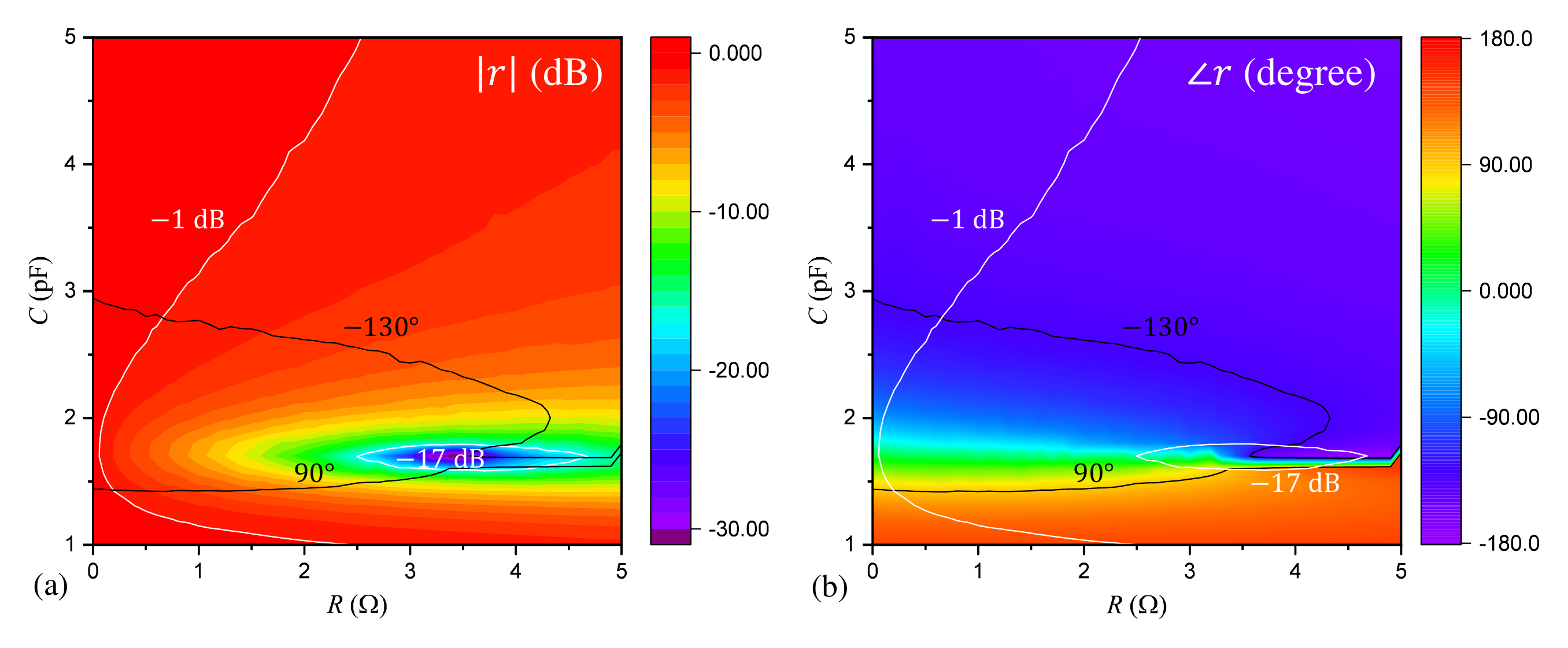}
	\caption{Independent control over the reflection amplitude in decibels (a) and phase in degree (b) by tuning the load resistance $R$ and capacitance $C$. The white and black curves show example contours of constant amplitude and constant phase respectively.}
	\label{fig:3}
\end{figure*}
As the intelligent metasurface is capable of tuning both the resistance and reactance, it can realize many tunable functions that cannot be achieved in conventional ways. For example, since perfect absorption takes place when the input impedance of the metasurface is matched to the free-space impedance, a change of the incident polarization and incident angle changes the required input impedance for TPA~\cite{Luukkonen2008}. Although the required input impedance remains purely resistive for any incidence, it is necessary to tune both surface resistance and reactance of the patch array. This tuning can be done by properly adjusting the properties of the metasurface globally and continuously, i.e., all unit cells must have the same configuration but the allowed settings must not be limited to a discrete set of values.

In the designed metasurface, the absorption coefficient of the metasurface is given by $A=1-\vert\Gamma\vert^2$ where $\vert\Gamma\vert$ denotes the absolute value of the reflection coefficient. Our goal is to achieve a very low value for the reflection coefficient ($\vert\Gamma\vert<-30$ dB) to obtain perfect absorption. We investigate the TPA feature for different incidence angles and two  main polarizations. The first one is the transverse-electric polarization where the electric field is in the $y$ direction (the incidence plane is the $x-z$ plane), and the second one is the transverse-magnetic polarization in which the magnetic field is parallel to the $x$ axis (the incidence plane is the $y-z$ plane), see Fig.~\ref{fig:MainSchem}. 

To find the required capacitance and resistance values provided by the ICs (series topology as shown in the inset of Fig.~\ref{fig:MainSchem}) corresponding to perfect absorption at 5~GHz, we follow a numerical approach and simulate the structure using ANSYS HFSS. For normal incidence ($\theta_{\rm{i}}=0^\circ$) in which two polarizations are degenerate, we find from simulations that we need $R=3.44~\Omega$ and $C=1.69$~pF to fully absorb the incoming wave. The corresponding reflection coefficient at 5~GHz is $|\Gamma|=-59.4$~dB indicating nearly perfect absorption. For oblique incidence, varying the incidence angle from $10^{\circ}$ to $70^{\circ}$ for both polarizations, we also obtain the required $R$ and $C$ values; they are summarized in Fig.~\ref{fig:2}. As we can observe, for the TE polarization, the required capacitance does not change dramatically, and it is approximately  $C\approx 1.68$~pF. However, the required resistance varies from 0.8~$\Omega$ (corresponding to $70^{\circ}$) to 3.4~$\Omega$ (corresponding to $10^{\circ}$). On the other hand, the scenario for the TM polarization is different. Both $R$ and $C$ experience noticeable changes as the incident angle varies. The different TE/TM behavior originates from the different impedance matching conditions for oblique incidence angles. First of all, let us consider the input impedance $Z_{\mathrm{inp}}$ for the grounded patch array metasurface, i.e., without the tunable ICs. In this case, when the incidence angle varies, the main difference of $Z_{\mathrm{inp}}$ is on the reactive parts \cite{Luukkonen2008} (see Fig.~S4 in Supplemental Material for details). With the increase of the incidence angle, while the reactive part of $Z_{\mathrm{inp}}$ decreases considerably for the TM polarization, it remains almost unchanged for TE polarization. This behavior means that if we want perfect absorption at larger incidence angles and at a fixed frequency, larger capacitance is required to compensate for the reactance reduction in TM polarization, but not in TE polarization. On the other hand, when varying the incidence angle $\theta$, the free-space impedance $Z_0$ also changes differently for the two polarizations, i.e., $Z_{\mathrm{0,TE}}=\eta_0/\cos\theta$ and $Z_{\mathrm{0,TM}}=\eta_0\cos\theta$. This result means that different resistance modulations are required for the two polarizations. The changes of the required resistance are also clearly seen in Fig.~S5 in Supplemental Material, where the color maps show the reflection amplitude (in decibels) while sweeping the $R$ and $C$ values of ICs, for the TE polarization and for three different incidence angles. In addition, Fig.~S5 quantifies the tolerances about the optimal values. Setting a limit, e.g. $|\Gamma|<-20$~dB, for achieving perfect absorption, we find that the tolerances in the required $R$ and $C$ values are ample, even at steep incidence angles.

The required $R$ and $C$ values summarized in Fig.~\ref{fig:2} are practical and they can be easily realized using electronic circuits~\cite{AdelS.Sedra2014}. More importantly, the values can be modified via bias voltages feeding the electronic circuits. Therefore, by continuously tuning $R$ and $C$ from one combination to another by varying bias voltages, we can achieve TPA for different polarizations and different incidence angles. It should be noted that we can also obtain TPA at different frequencies from 4.5~GHz to 5.5~GHz as the resonance frequency (where perfect absorption happens) can be easily tuned with $C$ (see Figs.~S1 and S6 in Supplemental Material for details).

Moreover, the unit cell design can be extended into an isotropic one, i.e., with $2\times 2$ patches where each pair of the neighboring patches is connected to a tunable IC (see Fig.~S7 in Supplemental Material for the schematic). Such an extended configuration has rotational symmetry and it is still nonmagnetic. Therefore, it will respond identically to both polarizations with the electric field along $y$ or $x$ directions, and the required $R$ and $C$ values depicted in Fig.~\ref{fig:2} are still valid.


\section{Tunable Anomalous Reflection}\label{sec:AnomRefl}
The independent control over $R$ and $C$ in the ICs enables us to freely and independently control the reflection amplitude and phase from the intelligent metasurface. As an illustration, Fig.~\ref{fig:3} shows the reflection amplitude and phase with respect to the changes of $R$ and $C$ when the metasurface is set homogeneously. As we can observe, the reflection amplitude can be modulated from $-30$ dB to almost 0 dB,  and for each value of the amplitude, the available reflection angle is quite broad. For example, if we fix the reflection amplitude to $-1$ or $-17$~dB, as shown by the white curves in the figure, the accessible reflection phase is from $-150^{\circ}$ to $+140^{\circ}$ by tuning the $R$ and $C$ values following the white curve. On the other hand, for a desired reflection phase, for example, $-130^{\circ}$ or $+90^{\circ}$, shown by the black curves, one can tune the reflection amplitude in a wide range. Such independent control over the reflection amplitude and phase can be helpful for holography and beam-shaping applications. Moreover, together with the independent biasing of each IC in the unit cells, this intelligent metasurface enables other practically important and more challenging tunable functionalities.

One prominent example is tunable perfect anomalous reflection as a means to manipulate wavefronts. Anomalous reflection, in which the impinging wave is deflected away from the specular direction \cite{Yu:2011,Diaz-Rubio:2017,Asadchy:2017,Wong2018,Radi2017,Kwon:2017PRB,Rabinovich2018,Chalabi2017}, attracted much attention since the realization with a linear reflection phase gradient according to the generalized Snell's law \cite{Yu:2011}. However, it was shown that in general  this approach results in parasitic reflections to undesired directions (e.g., \cite{Diaz-Rubio:2017}). In fact, these parasitics can be suppressed by constructing appropriately designed supercells that promote reflection to a specific diffraction order while suppressing radiation to other diffraction orders \cite{Diaz-Rubio:2017,Asadchy:2017,Wong2018,Radi2017,Kwon:2017PRB,Rabinovich2018}, realizing nearly perfect anomalous reflection. However, the existing realizations work only for a particular reflection angle and do not support tunable reflection angles. Moreover, for some ranges of the incidence and reflection angles, the metasurface must be strongly nonlocal, which has been realized only using dense arrays of patches of different, carefully optimized sizes. Here, enabled by the tunability of the ICs, we show that the intelligent metasurface can support tunable perfect anomalous reflection to the first and second diffraction orders, and for both TE and TM polarizations, although all unit cells have the same dimensions. To achieve these properties, we start from the prescription of a linear phase profile and subsequently use optimization to fine-tune the design.

\begin{figure}[t!]\centering
	\includegraphics{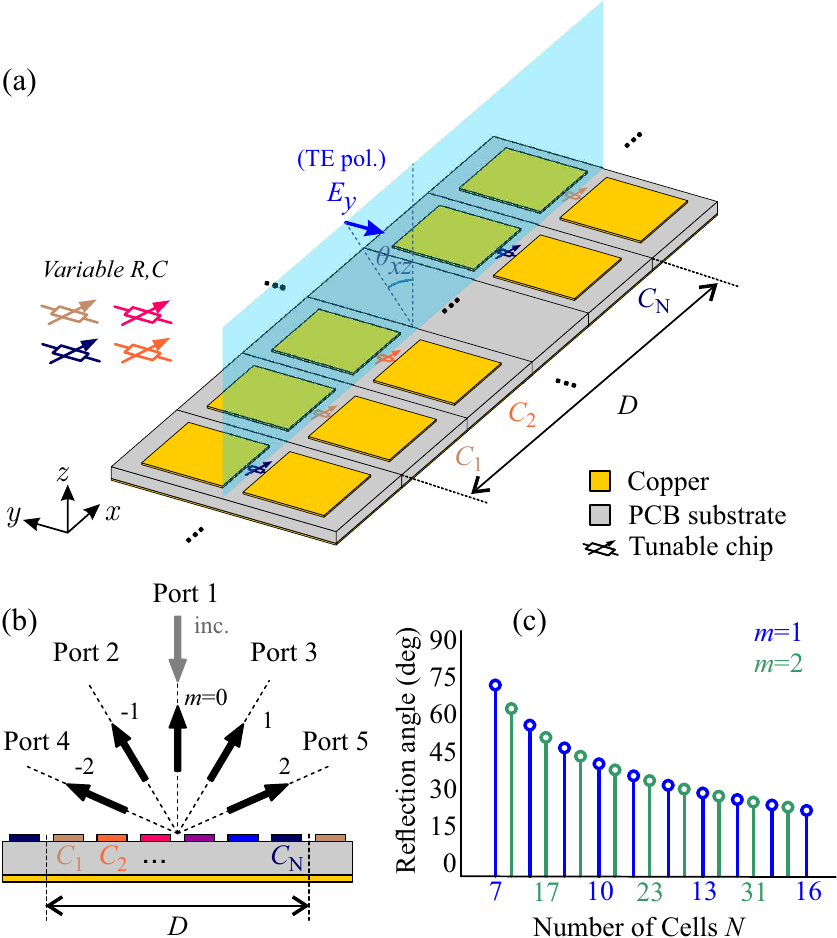}
	\caption{\label{fig:SuperSchem} (a)~Schematic view of the supercell for anomalous reflection in the $x-z$ plane. Different colors of the lumped elements represent different capacitance settings. (b)~Port naming convention and correspondence with reflected diffraction orders for normal incidence. (c)~Anomalous reflection angles for different numbers of cells exploiting the first and second diffraction orders (assuming normal incidence).}
\end{figure}


\subsection{Supercell description}
For achieving perfect anomalous reflection, we construct a supercell of $N$ unit cells (the total extent is $D=Nd$) with varying capacitances $C_1, C_2,\ldots,C_N$. In Fig.~\ref{fig:SuperSchem}(a) a supercell along the $x$ axis is depicted that can perform anomalous reflection in the $x-z$-incidence plane. Changing the supercell size enables anomalous reflection to different directions. For example, when the supercell size $D$ is $2\lambda_0 < D < 3\lambda_0$ and under normal incidence, four diffraction orders ($m=\pm1$, $\pm2$) besides the specular ($m=0$) are propagating; one port is assigned to each diffraction order for measuring the corresponding power and the naming convention is shown in Fig.~\ref{fig:SuperSchem}(b). The reflection angle for a given diffraction order, $m$, is dictated by momentum conservation $k_0(\sin\theta_r-\sin\theta_i)=m(2\pi/D)$, which yields
\begin{equation}
	\theta_r=\arcsin(m\lambda_0/D)
\end{equation}
for normal incidence ($\theta_i=0$). Note that with different sizes of the supercell, promoting different diffraction orders allows us to achieve a quasicontinuous coverage for the reflection angle. This feature is illustrated in Fig.~\ref{fig:SuperSchem}(c) where we plot the supported anomalous reflection angles for the first and second diffraction orders as the number of cells in the supercell varies.

\begin{figure}[t!]\centering
	\includegraphics{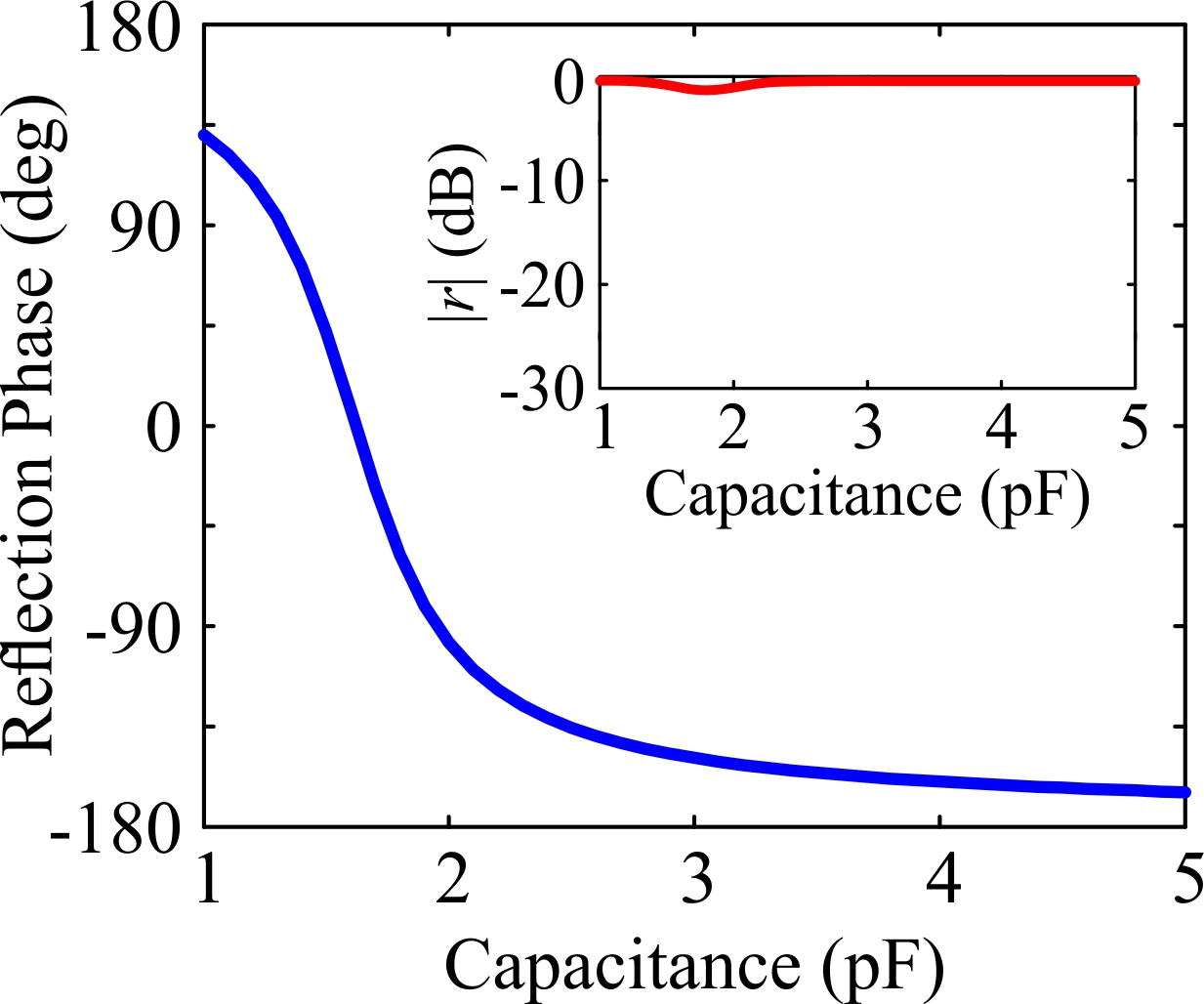}
	\caption{\label{fig:Lookup}Look-up table: Reflection phase for a uniform metasurface as a function of the capacitance of the tunable IC. The operating frequency is 5~GHz and the incident wave impinges at normal incidence. The corresponding reflection amplitude is shown in the inset.}
\end{figure}

As a first step in the design process, we impose a linear phase profile along the supercell $\phi(x)=\phi_0-m(2\pi/D)x$, thus promoting a specific diffraction order over the remaining leakage channels. We note that the resistance $R$ of the ICs is purposely set to zero for all unit cells to minimize absorption. Subsequently, we utilize optimization to fine-tune the design and achieve perfect anomalous reflection. In order to specify the required capacitances for the initial linear-phase prescription, we need a ``look-up table'' relating the reflection phase with the capacitance of the load. It is specified by illuminating the uniform metasurface with a normally incident plane wave at the operation frequency of 5~GHz, and the results are shown in Fig.~\ref{fig:Lookup}. To be realistic, we limit the achievable series capacitances in the range $[1, 5]$~pF. Even under this restriction, we have access to a large reflection phase span of $300^{\circ}$, while the reflection amplitude remains almost unity (inset).

\subsection{Transverse-electric polarization\label{sec:xzInc}}
\begin{figure}[h!]\centering
	\includegraphics{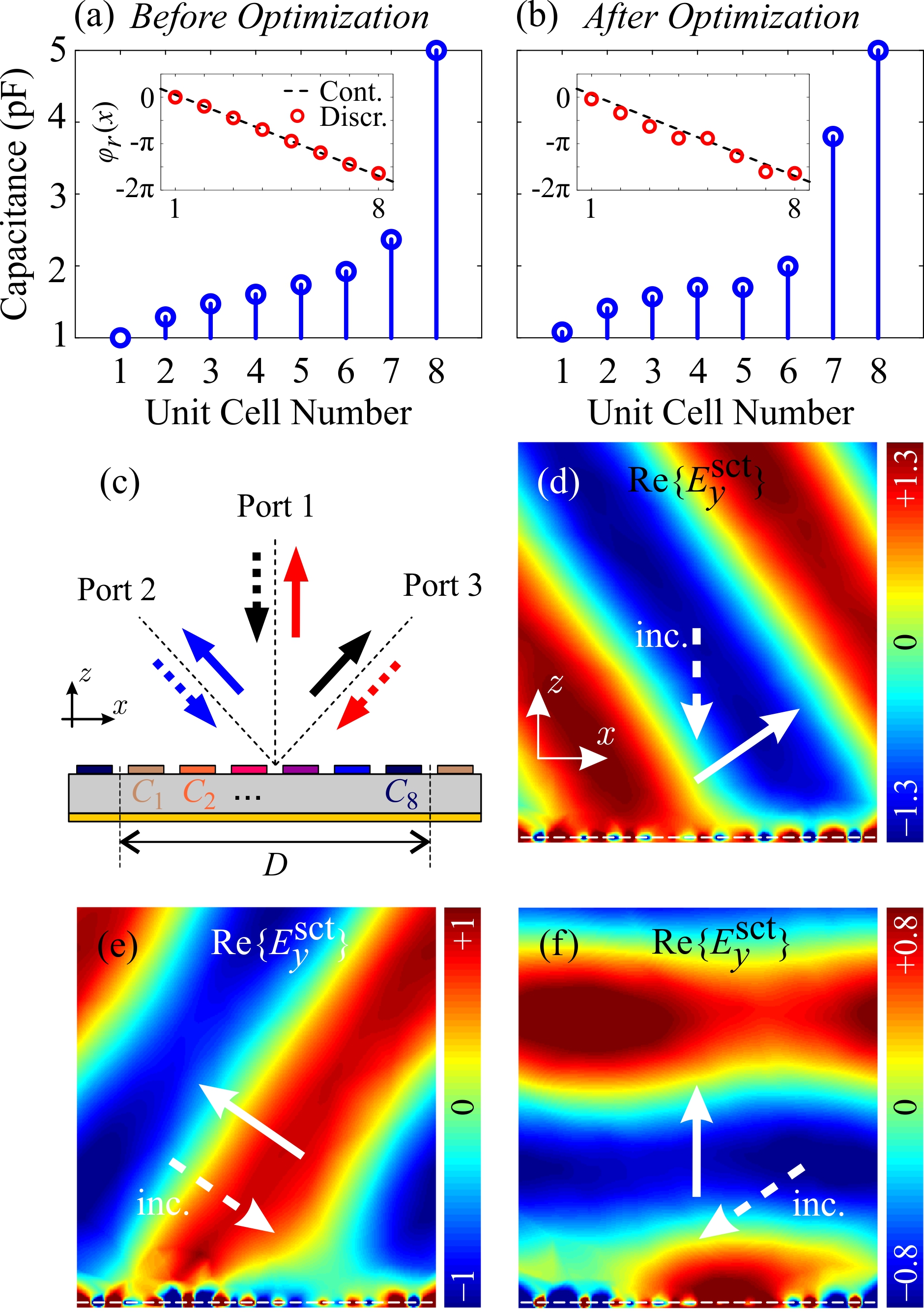}
	\caption{\label{fig:N8TE} Anomalous reflection with an $N=8$ supercell promoting the first diffraction order. (a),(b)~Capacitance values for the meta-atoms before optimization (linear phase prescription) and after optimization. The corresponding phase profiles are shown in insets. (c)~Port naming convention and the three reflection scenarios: $1\rightarrow3$, $2\rightarrow2$, $3\rightarrow1$. (d),(e),(f)~Scattered electric field for the three reflection scenarios. Planar wavefronts tilted to the desired direction are observed.}
\end{figure}

As a first example, we consider a supercell consisting of $N=8$ unit cells stacked along the $x$ axis $(D=Nd_x=72.96~\mathrm{mm}\sim1.2\lambda_0)$. For the normal incidence we get $m=\pm1$ diffraction orders in the $\pm 55.3^\circ$ directions. The supercell can be, thus, effectively described by a three-port network, as shown in Fig.~\ref{fig:N8TE}(c). Using the look-up table in Fig.~\ref{fig:Lookup} we specify the capacitances of the lumped loads for achieving a linear phase profile and promoting the $m=1$ diffraction order; they are depicted in Fig.~\ref{fig:N8TE}(a). The corresponding discrete reflection phases are shown in the inset. Notice that the first and last points deviate slightly from the prescription of a linear reflection phase. This fact occurs because we limit the available capacitance range to $[1,\ 5]$~pF.

Using the capacitance values given in Fig.~\ref{fig:N8TE}(a) we characterize the supercell response by successively exciting the three ports and extracting the reflection coefficients in each port ($S$ parameters). The entire $S$-parameter matrix is (absolute values of the components in decibels)
\begin{equation}\label{eq:SparamsN8TE}
	|S|_\mathrm{dB}^\mathrm{init.}=
	\begin{bmatrix}
		-26.65 &  -12.55 & -0.68 \\
		-12.55 &  -2.01 & -10.15 \\
		-0.68 &  -10.15 & -28.36 \\
	\end{bmatrix}.
\end{equation}
Notice that the matrix is symmetric as dictated by reciprocity. Upon normal incidence (port 1), the incident power is indeed mostly reflected to the $m=1$ diffraction order (port~3), as can be seen by the high amplitude of the $S_{31}$ element in Eq.~\eqref{eq:SparamsN8TE}. However, there is some unwanted radiation in port~2 ($m=-1$) as well $(\vert S_{21}\vert=-12.55~\mathrm{dB})$. Parasitic reflections are anticipated since (i)~the periodic approximation under which the look-up table of Fig.~\ref{fig:Lookup} is obtained is not exact when we switch from the unit cell to the supercell and (ii)~there is a nonzero variation of the local reflection amplitude along with the local reflection phase. These are known limitations of the ``phase-gradient'' approach and strategies to overcome them have been proposed in the literature \cite{Diaz-Rubio:2017,Asadchy:2017,Wong2018,Radi2017}. Here, we use optimization to achieve perfect anomalous reflection. Specifically, we seek  capacitances that maximize $|S_{31}|$ (the optimization goal is set to $|S_{31}|>-0.5$~dB), while at the same time minimizing both $|S_{21}|$ and $|S_{11}|$ (the optimization goal is set to $|S_{21}|,|S_{11}|<-20$~dB). The optimized capacitance values are plotted in Fig.~\ref{fig:N8TE}(b) and the obtained $S$-parameter matrix for the optimized metasurface reads:
\begin{equation}\label{eq:SparamsN8TEopt}
	|S|_\mathrm{dB}^\mathrm{opt.}=
	\begin{bmatrix}
		-20.17 &  -20.50 & -0.47 \\
		-20.50 &  -1.23 & -16.11 \\
		-0.47 &  -16.11 & -18.39 \\
	\end{bmatrix}.
\end{equation}
As we can observe, with the optimized supercell, we increase the reflection towards port~3 and suppress radiation towards port 2, thus achieving nearly perfect anomalous reflection. This result is lucidly illustrated in Fig.~\ref{fig:N8TE}(d) where we plot the scattered electric field: nicely planar wavefronts tilted toward port~3 are observed, with negligible parasitic reflections.

Although optimization is performed for normal incidence, we improve the metasurface performance for incidence from ports 2 and 3 as well. Excitation from port 3 leads to reflection to port 1 with a high amplitude of  $|S_{13}|=|S_{31}|=-0.47$~dB (reciprocity) and excitation from port 2 leads to retroreflection with a high amplitude of $|S_{22}|=-1.23$~dB. The above property can be clearly seen in Fig.~\ref{fig:N8TE}(e),(f) where the scattered electric field is depicted for the two cases showing minimal parasitic reflections. We note that the performance of this configuration is asymmetric in the angular direction as illuminations from $\pm55.3^\circ$ result in different reflection angles.

In order to further quantify the efficiency of the anomalous reflection we define the metric $\operatorname{Eff}_i=\max_{x}|S_{xi}|^2/\sum_{x}|S_{xi}|^2$, where the summation over $x$ runs through the number of ports. The values before and after optimization are $[97.7,\ 80.5,\ 89.7]\%$ and $[98.0,\ 95.8,\ 95.8]\%$, respectively, illustrating the improvement in performance for all excitation scenarios. Finally, we can calculate the absorption in the metasurface when illuminating from the three ports through $\operatorname{Abs}_i=1-\sum_{x}|S_{xi}|^2$ or directly from the electromagnetic field distribution by integrating the power loss density in patches, as well as in substrate and lumped loads. For the optimized supercell it equals $[8.47,\ 21.33,\ 6.42]\%$. Notice that the retroreflection operation results in increased absorption due to higher local field values on the metasurface.

With computer-controlled ICs we can easily rearrange the supercell to achieve anomalous reflection into different angles, allowing for reconfigurable operation. To better illustrate this concept we next study the case $N=9$ corresponding to $m=\pm1$ diffraction orders toward $\pm46.9^\circ$. We follow the same design procedure as with $N=8$. The $S$-parameter matrices before and after optimization are
\begin{gather} \label{eq:SparamsN9TE}
	|S|_\mathrm{dB}^\mathrm{init.}=
	\begin{bmatrix}
		-27.51 &  -15.02 & -0.47 \\
		-15.02 &  -1.82 & -12.80 \\
		-0.47 &  -12.80 & -27.09 \\
	\end{bmatrix},\\ \label{eq:SparamsN9TEopt}
	|S|_\mathrm{dB}^\mathrm{opt.}=
	\begin{bmatrix}
		-30.17 &  -31.73 & -0.32 \\
		-31.73 &  -1.36 & -21.66 \\
		-0.32 &  -21.66 & -27.43 \\
	\end{bmatrix}.
\end{gather}
The respective efficiencies for the three excitation scenarios are [96.43, 88.69, 94.28]\% and [99.82, 98.98, 99.08]\%, before and after optimization, respectively. As in the previous case, we obtain excellent performance for all three excitation scenarios despite specifically optimizing for incidence from port~1. The capacitance values for the initial linear phase prescription and the optimized supercell can be found in Table~\ref{tab:N9TE}.

\begin{table}[h!]\centering
	\begin{ruledtabular}
		\caption{\label{tab:N9TE}$N=9$ supercell for anomalous reflection at $\theta_r=46.9^{\circ}$ in the $xz$ incidence plane. Initial (linear phase profile) and optimized capacitance values.}
		\begin{tabular}{rccccccccc}
			$C$(pF)&$C_1$&$C_2$&$C_3$&$C_4$&$C_5$&$C_6$&$C_7$&$C_8$&$C_9$\\ \hline
			Init. &1.00&1.24&1.43&1.56&1.67&1.80&2.00&2.53&5.00\\
			Opt. &1.00&1.38&1.40&1.60&1.77&1.79&2.03&4.36&5.00\\
		\end{tabular}
	\end{ruledtabular}
\end{table}

\begin{figure}[t!]\centering
	\includegraphics{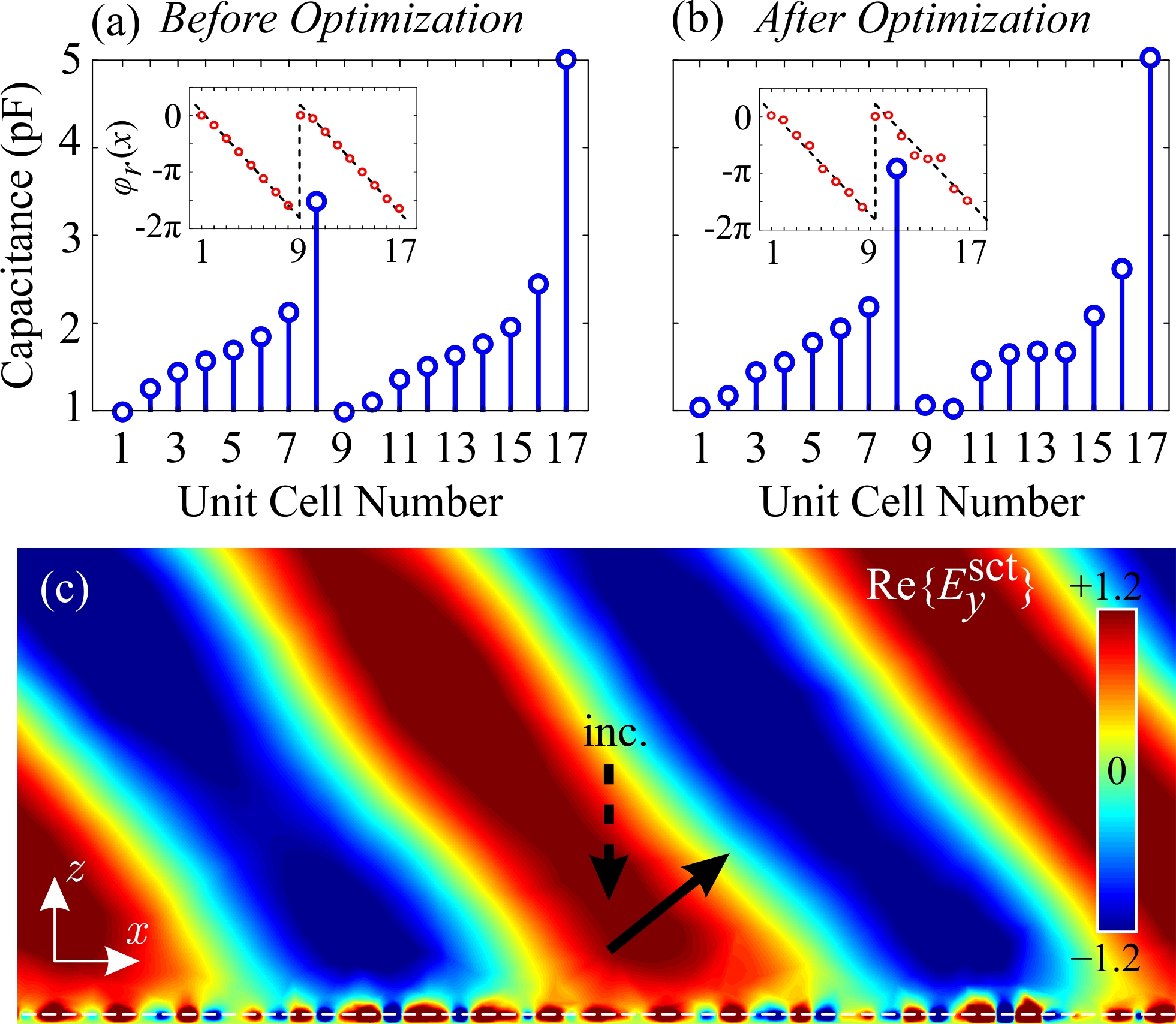}
	\caption{\label{fig:N17TE} Anomalous reflection with an $N=17$ supercell promoting the second diffraction order. (a),(b)~Capacitance values for the meta-atoms before optimization (linear phase prescription) and after optimization. The corresponding phase profiles are shown in the insets. (c)~Scattered electric field when illuminating from port~1: perfect anomalous reflection toward $50.7^\circ$ (port~5).}
\end{figure}

As shown in Fig.~\ref{fig:SuperSchem}(c), utilizing higher diffraction orders, we can bridge the gap between the discrete reflection angles resulting from operation in the first diffraction order. Next, we choose $N=17$ for steering toward $50.7^\circ$ ($m=2$), which lies in between $55.3^\circ$ and $46.9^\circ$ of the two previous examples. In this case, the linear phase profile used as an initial prescription covers the $0-4\pi$ span [inset in Fig.~\ref{fig:N17TE}(a)]. The initial and optimized capacitances are depicted in Fig.~\ref{fig:N17TE}(a) and (b), respectively. The corresponding $S$ parameters before and after optimization for excitation from port~1 are $|S_{x1}|_\mathrm{dB}^\mathrm{init.}=$[-21.50 -13.50 -27.00 -37.02 -0.60] and $|S_{x1}|_\mathrm{dB}^\mathrm{opt.}=$[-20.75 -21.56 -21.87 -20.09 -0.51]. After optimization, the unwanted radiation to port~2 ($m=-1$ diffraction order) is suppressed and normally incident illumination is almost exclusively reflected into the $m=2$ diffraction order with an efficiency of 96.5\%. This result can also be seen in Fig.~\ref{fig:N17TE}(c) where we plot the scattered electric field.


\subsection{Transverse-magnetic polarization\label{sec:yzInc}}
\begin{figure}[t!]\centering
	\includegraphics{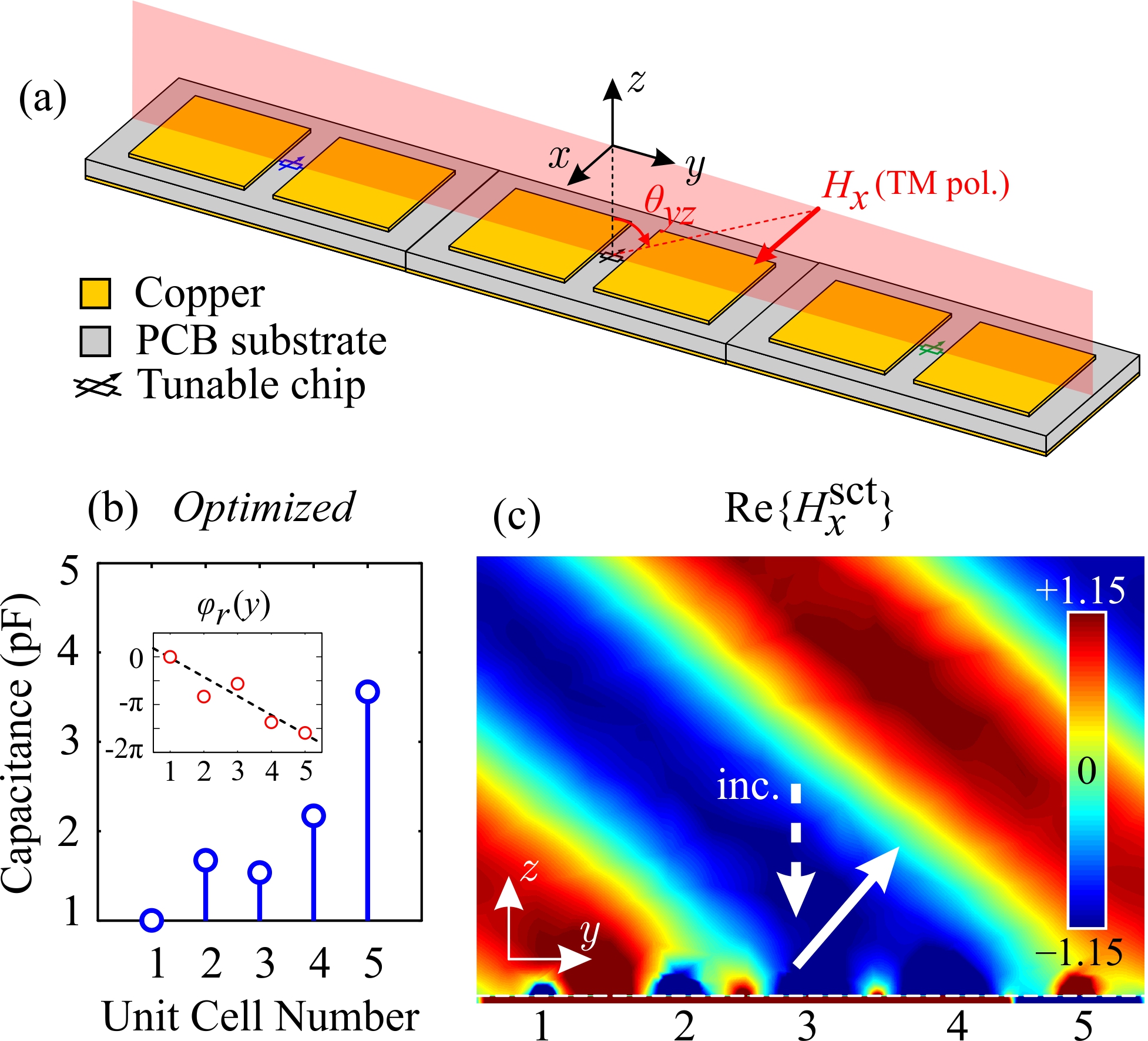}
	\caption{\label{fig:N5TM} Anomalous reflection in the $y-z$ plane (TM polarization) with a $N=5$ supercell along the $y$ axis promoting the first diffraction order. (a)~Schematic of unit-cell stacking along the $y$ axis with $N=3$. (b)~Capacitance values for the meta-atoms after optimization; the corresponding phase profile is shown in the inset along with the linear phase profile prescription (dashed line). (c)~Scattered magnetic field when excited from port~1: perfect anomalous reflection toward $41.1^\circ$ (port 3).}
\end{figure}

The unit cells can be readily rearranged by varying the tunable impedances of the unit-cell loads. This property enables us to get perfect anomalous reflection also for the TM polarization ($\mathbf{H}=H_x\hat{\mathbf{x}}$) in the $y-z$ plane. In this case, we construct supercells along the $y$ axis [Fig.~\ref{fig:N5TM}(a)] to realize the required phase gradient along the $y$ direction. We consider the case of $N=5$ $(D=Nd_y=91.2~\mathrm{mm}\sim1.5\lambda_0)$ giving rise to $\pm1$ diffraction orders toward $\pm41.1^{\circ}$.

Again, we start with the prescription of a linear phase profile and subsequently use optimization to fine-tune the load capacitances and the corresponding phase distribution [Fig.~\ref{fig:N5TM}(b)]. The resulting $S$-parameter matrices before and after optimization are
\begin{gather} \label{eq:SparamsN5TM}
	|S|_\mathrm{dB}^\mathrm{init.}=
	\begin{bmatrix}
		-15.72 &  -9.18 & -1.02 \\
		-9.18 &  -2.01 & -7.70 \\
		-1.02 &  -7.70 & -34.66 \\
	\end{bmatrix},\\ \label{eq:SparamsN5TMopt}
	|S|_\mathrm{dB}^\mathrm{opt.}=
	\begin{bmatrix}
		-27.66 &  -25.48 & -0.32 \\
		-25.48 &  -0.65 & -20.33 \\
		-0.32 &  -20.33 & -25.70 \\
	\end{bmatrix}.
\end{gather}
The respective efficiencies for the three excitation scenarios are [84.27,  69.42,  82.29]\% before and [99.51, 98.61, 98.73]\% after optimization, respectively. Compared to the examples in Sect.~\ref{sec:xzInc}, the initial efficiencies are not as high. This reduction is because the unit cell is not as subwavelength in the $y$ direction ($d_y=2d_x=18.24~\mathrm{mm}\sim\lambda_0/3$), leading to a coarser discretization of the required phase profile [Fig.~\ref{fig:N5TM}(b)]. However, even in this case, excellent performance is obtained through optimization. The scattered magnetic field when the metasurface is illuminated from port~1 is depicted in Fig.~\ref{fig:N5TM}(c), showing perfect anomalous reflection toward $41.1^\circ$ (port 3) in the $y-z$ plane.


\section{Conclusions}\label{sec:concl}
In summary, we design and numerically evaluate a programmable, intelligent metasurface capable of tunable perfect absorption and tunable perfect anomalous reflection in the microwave band. This dual functionality is enabled by equipping the unit cells with locally tunable ICs that provide a continuously tunable complex impedance, in \emph{both} the resistive and the reactive parts. Compared to conventional tunable metasurface designs relying on switch diodes (discrete binary load values) or varactors (tunability only in the reactive part of the loads), the introduction of the fully controllable IC concept provides a much broader functionality spectrum and wider tunability ranges with respect to the operating frequency and the incidence angle. The simulation results unveil the multifunctional and wide tunability features of this intelligent metasurface. It is noted that the unit cell design may not be optimal and better performance may be achieved with further optimizations. We stress that the presented results show only some example functionalities: with proper selection of the complex tunable loads, other functions can be provided, such as energy focusing at desired points, arbitrary wavefront shaping operations, reflection-mode holography, and more. This feature is enabled by the broad tunable span of reflection directions and a wide range of tunable absorption. The results show that, by properly tuning the bulk loads of individual cells, fundamental limitations of the designs based on controlling the local reflection phase can be overcome. Moreover, with arbitrary space and time modulation of the loads, our intelligent metasurface can open new opportunities for electromagnetic wave manipulation~\cite{Zhang2018,Mirmoosa2019}. We anticipate that this work will encourage and facilitate incorporation of artificial intelligence into metasurfaces for a wide variety of applications.

Further capitalizing upon the presence of the ICs in meta-atoms, we can envision that the unit cells are interconnected, forming an adaptive and concurrently reconfigurable network, thus paving the way toward fully programmable metasurfaces based on surface-distributed computing. Extra steps regarding the physical realization require further investigation, especially concerning the ICs which are the key enabling elements in this design.


\section*{Acknowledgments}
This work was supported by the European Union's Horizon 2020 Future Emerging Technologies call (FETOPEN-RIA) under Grant Agreement No.~736876 (project
VISORSURF). Work at FORTH was partially
supported by the European Research Council under the ERC
Advanced Grant No.~320081 (PHOTOMETA). Work at Ames Laboratory was supported by the US Department of Energy (Basic Energy Science, Division of
Materials Sciences and Engineering) under Contract No.~DE-AC02-
07CH11358.

\bibliography{Metasurfaces}

\end{document}